\documentclass[final,5p]{elsarticle}
\usepackage{lineno,hyperref}
\journal{Acta Materialia}
\usepackage{gensymb}
\usepackage{amsfonts}
\usepackage{siunitx}
\usepackage{amsmath}
\usepackage{amssymb}
\usepackage{graphicx}
\usepackage{dcolumn}
\usepackage{euscript}
\usepackage{multirow}
\usepackage{color}
\usepackage{dblfloatfix}

\providecommand{\U}[1]{\protect\rule{.1in}{.1in}}

\biboptions{sort&compress} 

\newcommand{\wlfig}{0.42\textwidth}
\newcommand{\rFig}[1]{Figure~\ref{#1}}
\newcommand{\rfig}[1]{Fig.~\ref{#1}}
\newcommand{\rtbl}[1]{Table~\ref{#1}}

\newcommand{\ms}{M_\text{s}}
\newcommand\tc{T_\text{C}}

\newcommand{\hci}{H_\text{ci}}

\newcommand{\br}{B_\text{r}}
\newcommand{\ala}{\alpha_1}
\newcommand{\alb}{\alpha_2}

\usepackage{float}
\newcommand{\atpercent}{at.\percent}

\begin{document}
\begin{frontmatter}
  
\title{On spinodal decomposition in alnico---a transmission electron microscopy and atom probe tomography study}

\author[ames]{Lin Zhou\corref{mycorrespondingauthor}}
\cortext[mycorrespondingauthor]{Corresponding author}
\ead{linzhou@ameslab.gov}
\author[oak]{Wei Guo}
\author[oak]{Jonathan D. Poplawsky}
\author[ames]{Liqin Ke}
\author[ames]{Wei Tang}
\author[ames]{Iver E. Anderson}
\author[ames]{Matthew J. Kramer}
\address[ames]{Ames Laboratory, U.S. Department of Energy, Ames, Iowa 50011, USA}
\address[oak]{Center for Nanophase Materials Sciences, Oak Ridge National Laboratory, Oak Ridge, TN 37831, USA}

\begin{abstract}
Alnico is a prime example of a finely tuned nanostructure whose
magnetic properties are intimately connected to magnetic annealing
(MA) during spinodal transformation and subsequent lower temperature
annealing (draw) cycles. Using a combination of transmission electron
microscopy and atom probe tomography, we show how these critical
processing steps affect the local composition and nanostructure
evolution with impact on magnetic properties. The nearly 2-fold
increase of intrinsic coercivity ($\hci$) during the draw cycle is not
adequately explained by chemical refinement of the spinodal
phases. Instead, increased Fe-Co phase ($\ala$) isolation, development
of Cu-rich spheres/rods/blades and additional $\ala$ rod precipitation
that occurs during the MA and draw, likely play a key role in $\hci$
enhancement. Chemical ordering of the Al-Ni-phase ($\alb$) and
formation of Ni-rich ($\alpha_3$) may also contribute. Unraveling of
the subtle effect of these nano-scaled features is crucial to
understanding on how to improve shape anisotropy in alnico magnets.
\end{abstract}

\end{frontmatter}


\section{Introduction}
Using an external magnetic field during heat-treatment, magnetic
annealing (MA), to control materials' microstructure has been widely
studied since the 1950s~\cite{mccurrie1982coll-c3sp}. MA promotes
formation of texture, biases spinodally decomposed phase morphologies,
promotes martensitic transformation in ferrous alloys, changes phase
transformation temperatures,
etc~\cite{watanabe2006sm,watanabe2006jms}. Cahn showed that MA is most
effective in alnico magnets at a temperature near the onset of
spinodal decomposition (SD) and below the Curie
temperature~\cite{cahn1963jap}. Alnico has recently re-attracted a
large amount of interest as a near-term, non-rare-earth permanent
magnetic alloy for wind power generators and electric vehicle
motors~\cite{kramer2012jjmm,zhou2014am,mccallum2014armr}. Its lower
cost and stable performance over a wide temperature range make it
still irreplaceable after eighty years of development.

Unlike rare-earth based permanent magnets, coercivity in alnico is
provided by shape anisotropy, instead of intrinsic magneto-crystalline
anisotropy. As a result, magnetic properties of alnico strongly depend
on the details of its unique microstructure: a periodically arrayed
elongated Fe-Co rich ($\ala$) hard magnetic phase embedded in a
continuous non-magnetic Ni-Al-rich ($\alb$) matrix, formed via
SD. Achieving the optimum properties in alnico requires a well
controlled and lengthy heat-treatment process, including
solutionization of the alloy above \SI{1200}{\celsius}, isothermal MA
near its Curie temperature and subsequent lower temperature annealing
(draw
cycles)~\cite{mccurrie1982coll-c3sp,stanek2010amm,sergeyev1970mito,takeuchi1976tjim,iwama1974tjim,iwama1970tjim}.
The application of MA marks the most important cornerstone in alnico
magnets' development history. MA biases the $\ala$ phase's morphology
during SD and make it grow along the $\langle100\rangle$
crystallographic direction closest to the external
field~\cite{zhou2014mmte}. The resulting anisotropic spinodal
nano-structure has significantly improved coercivity ($\hci$) and
remanence ($\br$) of alnico. The biased growth is optimal only when it
is performed within a narrow temperature range for a limited time. For
example, the ideal morphology (in transverse cross-section) that
brings optimum $\hci$ in higher grade alnico 8 and 9 series is a
mosaic structure consisting of periodically arrayed $\sim$\SI{40}{nm}
diameter $\ala$ phases embedded in a continuous $\alb$ matrix obtained
by MA at $\sim$\SI{840}{\celsius}~\cite{zhou2017am}.

On the other hand, the microstructure and chemistry changes during the
lower temperature draw process are much more subtle, although draw
cycles play an equally important role in increasing $\hci$ in
alnico. For example, our recent study showed that for alnico 8H, its
coercivity was nearly doubled after drawing, compared to the MA
alone~\cite{zhou2017am}. More interestingly, the absolute increase
during the draw-related $\hci$ enhancement was almost independent of
the preceding MA temperature. However, the mechanism behind $\hci$
enhancement during draw remains surprisingly elusive and is not well
understood despite the fact that drawing has been practiced for
decades. Most previous understanding on draw enhancement of $\hci$ was
based on a simplistic assumption: lower temperature annealing results
in a larger composition separation and a larger magnetization
difference between $\ala$ and $\alb$ phases, which increases the
$\hci$~\cite{mccurrie1982coll-c3sp}. This assumption appears
oversimplified, especially considering the recent findings which
reveal an insignificant compositional variation before and after the
draw~\cite{zhou2017am}. This lack of understanding is mainly due to
our inability to access and evaluate the very subtle nature of
chemical and microstructural evolution during the draw. Revealing
these subtle effects appears crucial to improve shape anisotropy in
permanent magnets or reduce coercivity in soft magnetic alloys.

This study was designed to elucidate the structural and chemical
evolution in alnico at different stages during heat treatment,
especially during the draw process, as well as their relationship with
magnetic properties. An isotropic (alnico 8H)
32.4Fe-38.1Co-12.9Ni-7.3Al-6.4Ti-3.0Cu (wt\percent) alloy was chosen
for this investigation. A combination of electron backscatter
diffraction (EBSD), atom probe tomography (APT), and transmission
electron microscopy (TEM) techniques were used to better discover and
more precisely characterize the morphology and chemistry in SD
phases. Based on our comprehensive characterization results, we
discuss $\hci$ enhancement mechanisms beyond the conventional
explanation.

\section{Experimental details}
\subsection{Sample preparation and property measurement}
Batch of pre-alloyed powder was made by gas-atomization in Ames
Laboratory. Details on magnet alloy consolidation to full density by
hot isostatic pressing have been reported
elsewhere~\cite{tang2015itom}. The resultant alloy was polycrystalline
with randomly oriented grains.  Center sections of the alloy were cut
into \SI{3}{\mm} diameter by \SI{8}{\mm} cylinders. The cylindrical
samples were solutionized at \SI{1250}{\celsius} for \SI{30}{\minute}
in vacuum and quenched in an oil bath (sample 1). Some samples were
then annealed at \SI{840}{\celsius} with an external applied field
of \SI{1}{\tesla} for \SIlist{0.5;1.5;5;10}{\minute}, with
corresponding samples labeled as 2 thru 5, respectively. This MA
temperature was determined from our previous study which gives optimum
alloy magnetic properties~\cite{tang2015itom}. Some samples also
underwent an additional low temperature drawing process
at \SI{650}{\celsius} for \SIlist{1;3;5}{\hour} (labeled 6 thru 8),
respectively. After MA or low temperature draw, the samples were water
quenched to room-temperature. Details of heat-treatment conditions of
samples 1--8 are listed in ~\rtbl{tbl:ht-condition}. Their magnetic
properties were measured using a Laboratorio Elettrofisico Engineering
Walker LDJ Scientific AMH-5 Hysteresis graph in a closed-loop setup.
\begin{table}[!hbtp]
  \footnotesize
  \centering
  \caption{Detailed heat-treatment conditions for samples 1 to 8.}
\label{tbl:ht-condition}
\bgroup
\def\arraystretch{1.2}
\begin{tabular}[0.9\linewidth]{cccccccccc}
  \hline
  \hline
  \\[-1.1em]
  {Samples}  & 1 & 2 & 3  & 4 & 5 & 6 & 7 & 8  \\
\\[-1.1em]
\hline
\\[-1.2em]
 MA/\SI{840}{\celsius}   (\si{\minute}) & 0 & 0.5 & 1.5  & 5 & 10 & 10 & 10 & 10  \\
 Draw/\SI{650}{\celsius} (\si{\hour})   & 0 & 0   & 0    & 0 &  0 &  1 &  3 &  5  \\
\hline
\end{tabular}
\egroup
\end{table}

\subsection{Characterization}
\begin{figure}[!thb]
\centering \includegraphics[width=0.41\textwidth,clip]{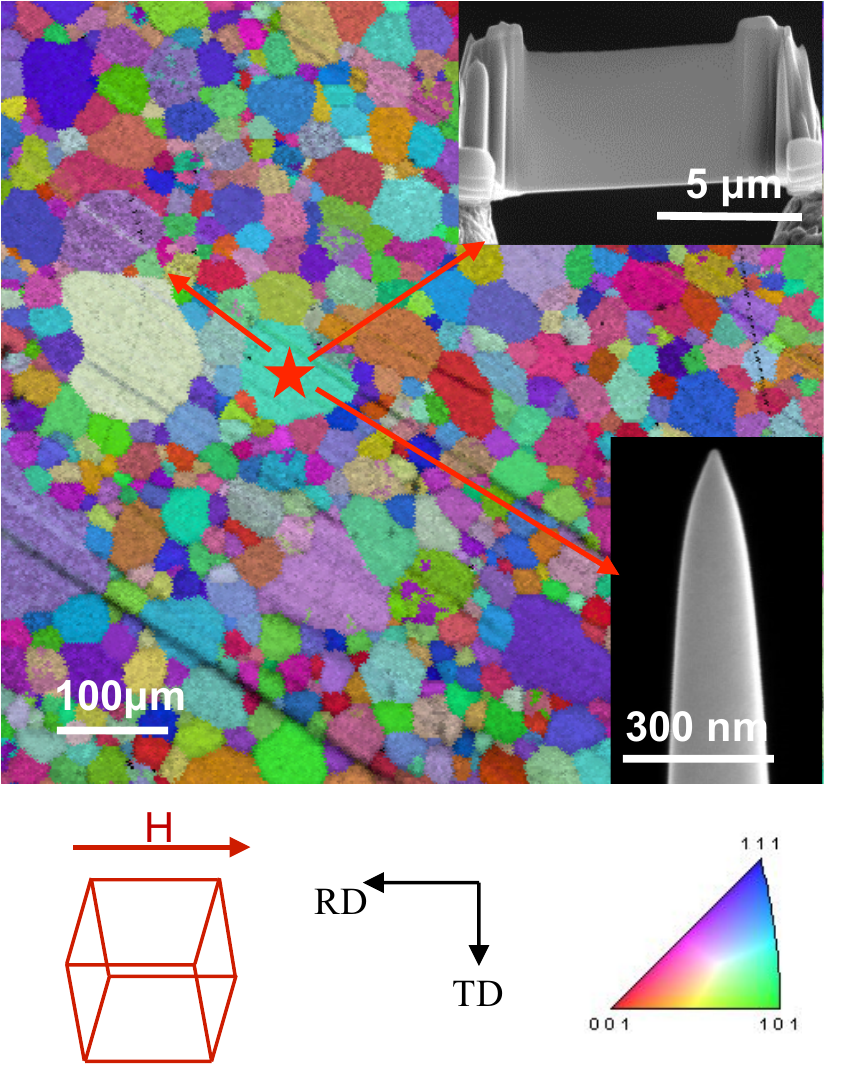}%
\caption{EBSD inverse pole figure of sample 5. Insets show a TEM (top) and APT (bottom) sample lifted from the same grain.}
\label{fig:01}
\end{figure}
APT and TEM are sensitive methods for investigating details of
structure and compositions of materials down to atomic-scale
resolution~\cite{miller2007sia,williams1996book}. In this study, APT
analysis of samples 1 to 8 was performed to better understand the SD
phase morphology in three dimensions, as well as a more accurate
determination of the $\ala$ and $\alb$ composition and morphology at
different heat-treatment stages. TEM was used to reveal the
crystallographic relationship between different phases. In each
sample, grains with their $\langle001\rangle$ crystal orientation
parallel to the external magnetic field direction were first
identified on the polished longitudinal sections (electron beam
perpendicular to external field direction) using EBSD, on an Amray
1845 field emission SEM. This can exclude the effect of orientation
difference between crystallographic $\langle001\rangle$ and external
field on SD microstructure~\cite{zhou2014mmte}. Both APT and TEM
samples were then lifted-out from the same selected grain to proceed
with a more detailed structural analysis, as shown in \rfig{fig:01}.
An FEI Nova 200 dual-beam focused ion beam (FIB) instrument was used
to perform lift-outs and annular milling of targeted grains to
fabricate needle-shaped APT specimens. A wedge lift-out geometry was
used to mount multiple samples on a Si microtip array to enable the
fabrication of several needles from one wedge
lift-out~\cite{thompson2007u}. APT was performed with a local
electrode atom probe (LEAP) 4000X HR manufactured by CAMECA
Instruments. Samples were run in voltage mode with a base temperature
of \SI{40}{\kelvin} and \SI{20}{\percent} pulse fraction at a
repetition rate of \SI{200}{\kHz}. The datasets were reconstructed and
analyzed using the IVAS 3.6.12 software (CAMECA Instruments). An FEI
Helios dual-beam FIB was used to perform lift-out of TEM samples with
the sample surface normal parallel to the external field direction
(transverse direction). An FEI probe aberration corrected Titan Themis
TEM with a Super-X energy dispersive X-ray spectrometer (EDS) was used
for structural characterization.

\section{Experimental results}
\subsection{Magnetic properties}
\begin{figure}[!htb]
\centering
\includegraphics[width=0.4\textwidth,clip]{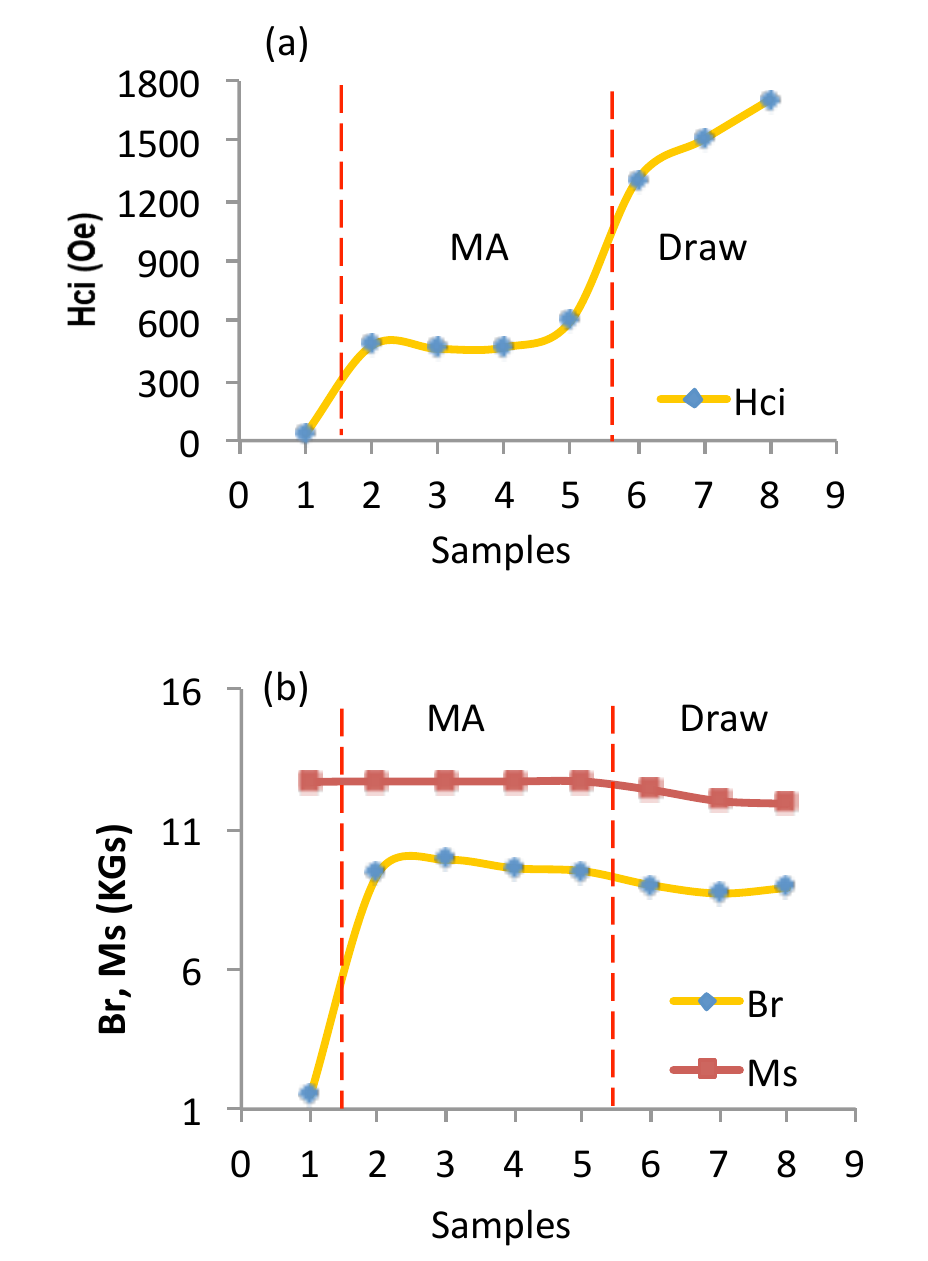}%
\caption{(a) Intrinsic coercivity ($\hci$), (b) remanence ($\br$), and saturation magnetization ($\ms$) of sample 1 thru 8.}
\label{fig:02}
\end{figure}

\begin{figure}[htb]
\centering
\includegraphics[width=0.49\textwidth,clip]{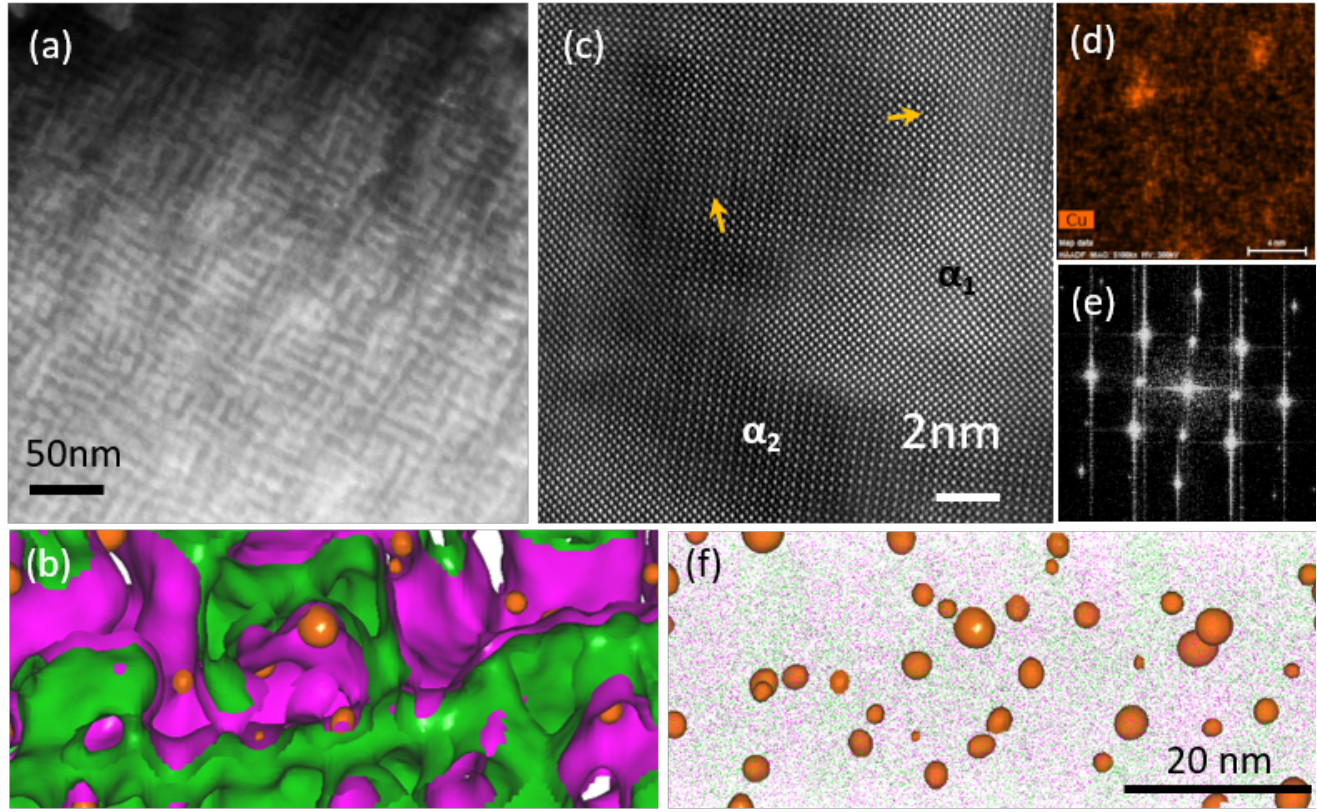}%
\caption{Sample 1 (a) HAADF STEM image showing internal SD.
(b) Three-dimensional atom probe data of the as-solutionized sample
(Sample 1) with isoconcentration surfaces clearly depicting the actual
nanostructure of the alloy. Isoconcentration values
are \SI{7.5}{\atpercent} Ni (green), \SI{10}{\atpercent} Cu (orange)
and \SI{30}{\atpercent} Fe (pink), which border the Ni, Cu, and Fe
rich regions, respectively. (c) HRSTEM image shows a coherent lattice
between $\ala$, $\alb$ and Cu-enriched clusters. Yellow arrows
indicate locations of Cu-enriched clusters. (d) and (e) are
corresponding EDS Cu mapping and FFT of (c), respectively. (f) spatial
distribution of Cu clusters, which are displayed
by \SI{10}{\atpercent} Cu isosurfaces.  Fe and Ni atoms are also
displayed.}
\label{fig:03}
\end{figure}

Magnetic properties of Samples 1-8 are summarized
in \rfig{fig:02}. Compared with the as-solutionized sample (sample 1),
short time MA (30 s, sample 2) already results in a dramatic
improvement of $\hci$ from \SI{35}{Oe} to \SI{476}{Oe}, and $\br$ from
\SI{1.5}{KGs} to \SI{9.4}{KGs}. Increasing the MA time to \SI{10}{min} further increase
$\hci$ ($\sim$\SI{26}{\percent}), but $\br$ plateaus
between \SI{30}{\second} and
\SI{10}{\minute} of MA. The saturation magnetization ($\ms$) of sample 1 and MA
samples (samples 2 to 5) is similar, indicating a similar volume
fraction of the $\ala$ phase in those samples~\cite{zhou2014am}.
Drawing can triple the $\hci$ to \SI{1707}{Oe} (sample 8), with the
most obvious improvement occurring after the initial drawing (Sample
6); however, both $\ms$ and $\br$ slightly drop after the draw.

\subsection{Microstructure after solutionization}

\begin{figure}[htb]
\centering
\includegraphics[width=\wlfig,clip]{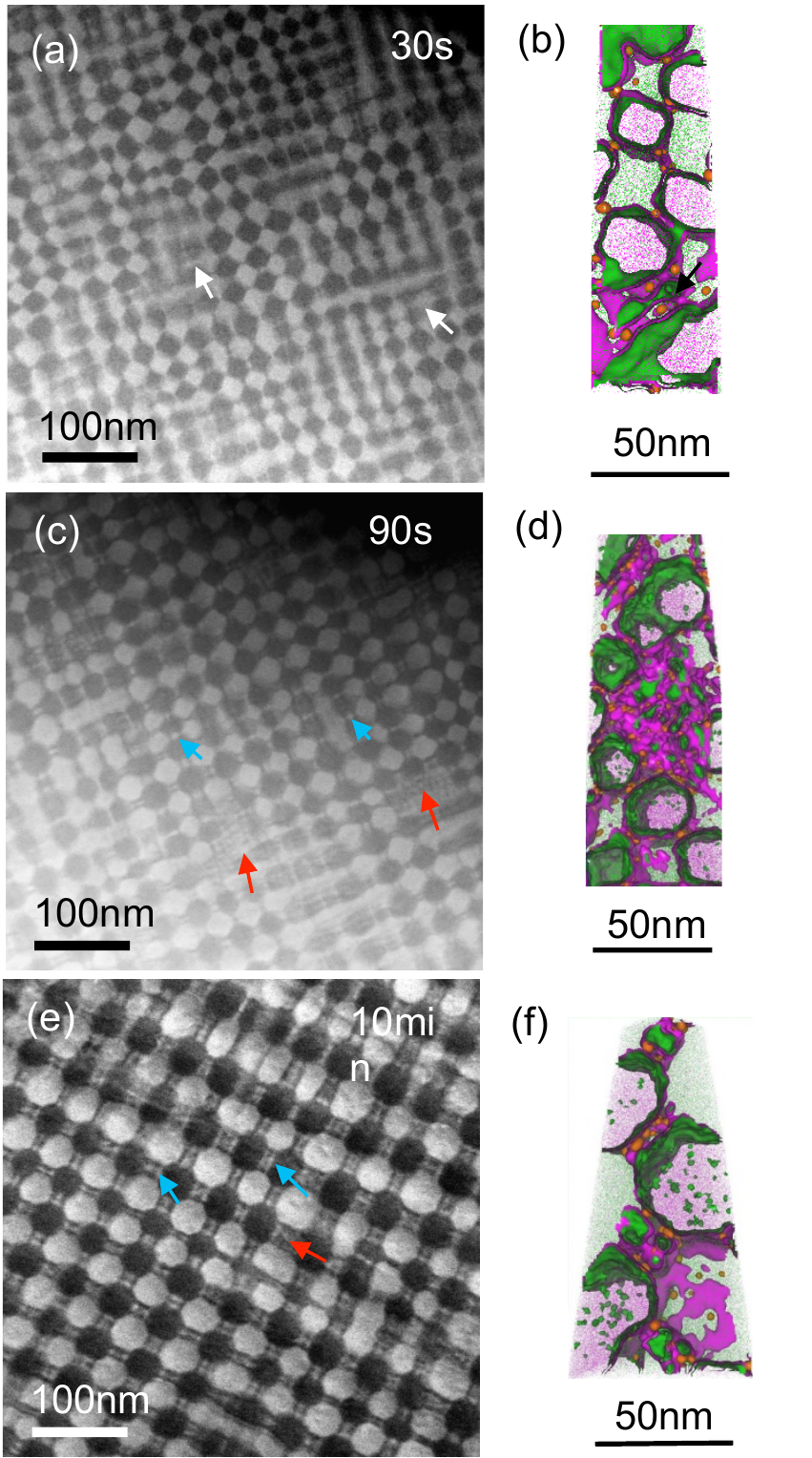}%
\caption{Nanostructures of sample 2 (a,b), sample 3 (c,d) and
sample 5 (e,f) revealed by HAADF STEM images and APT. Examples of
small $\ala$ phases between two adjacent large $\ala$ rods, and
regions with $\ala$ clusters are indicated by blue and red arrows,
respectively. White arrows indicate regions with a structure similar
to sample 1. For clarity, the APT reconstruction only shows the Ni
(green) and Fe (pink) atoms. Isoconcentration surfaces shown here are
\SI{7.5}{\atpercent} Ni (green), \SI{10}{\atpercent} Cu (orange) and \SI{30}{\atpercent} Fe (pink).  }
\label{fig:04}
\end{figure}

Chemical segregation has already begun after \SI{1250}{\celsius} of
solutionization, although the sample has almost no
coercivity. High-angle-annular-dark-field (HAADF) scanning
transmission electron microscopy (STEM) imaging was used to minimize
strain contrast and differentiate phase morphology more clearly than
can be achieved with traditional diffraction contrast TEM. The $\ala$
phase shows brighter contrast in HAADF STEM image due to the higher
averaged atomic number of the elements. A mixture of
$\sim$\SIrange{5}{10}{\nm} $\ala$ disks, and
$\sim$\SIrange{10}{40}{\nm} long rods with a diameter of
$\sim$\SIrange{5}{10}{\nm} was observed in sample 1
(\rfig{fig:03}a). The disks and rods are sometimes connected to each
other. The $\ala$/$\alb$ interface is slightly blurry, which may be
due to incomplete phase separation or sample thickness with respect to
the 3D microstructure. Isoconcentration surfaces within the APT data
of sample 1 reveals an interpenetrating nature of the $\ala$ and
$\alb$ phases, as shown in \rfig{fig:03}b. The $\ala$ and $\alb$
phases are continuous with meandering boundaries within the entire
analyzed volume. Both the $\ala$ and $\alb$ phases have
$\sim$\SIrange{5}{10}{\nm} diameters. The appearance of disk and rods
within the STEM images is a projected view of the $\ala$ phase along
different crystallographic directions.

A high density of Cu-enriched clusters were detected inside $\alb$, as
shown in \rfig{fig:03} b and f. The Cu-enriched clusters have an
average diameter of $\sim$\SI{1.2}{\nm} and occupy
$\sim$\SI{1.1}{\percent} of the alloy's volume. The Cu concentration
at the cluster center was measured to be $\sim$\SI{53.4}{\atpercent}
for sample 1 by APT. High-resolution HAADF STEM image
(HRSTEM, \rfig{fig:03}c) and corresponding fast-Fourier-transform
(FFT) (\rfig{fig:03}e) shows coherent $\ala$/$\alb$, $\alb$/Cu
interfaces. This result implies that the Cu-cluster and $\alb$ has the
same lattice structure. Moreover, due to small size of the
Cu-clusters, their positions (as indicated by arrows
in \rfig{fig:03}c) in the HAADF STEM image can only be identified by
overlaying the matching EDS elemental mapping (\rfig{fig:03}d).

\subsection{Microstructure after magnetic field annealing}

\begin{figure}[h]
\centering
\includegraphics[width=\wlfig,clip]{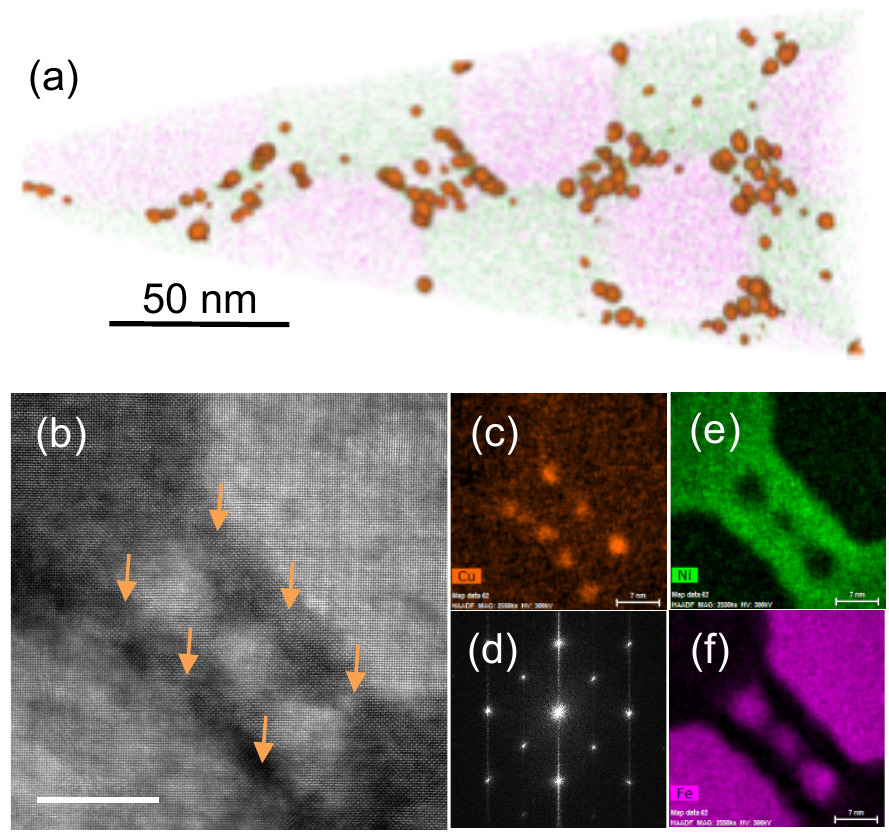}%
\caption{Sample 5 (a) APT reconstructed volumes show the
distribution of Cu clusters shown by \SI{10}{\atpercent}
isoconcentration surfaces, (b) HRSTEM image and corresponding (c), (e)
and (f) EDS Cu, Ni and Fe map, respectively (scale bar
is \SI{7}{\nm}). (d) FFT of (b). Orange arrows indicate location of
Cu-enriched clusters.  }
\label{fig:05}
\end{figure}

\begin{figure}[h]
\centering
\includegraphics[width=\wlfig,clip]{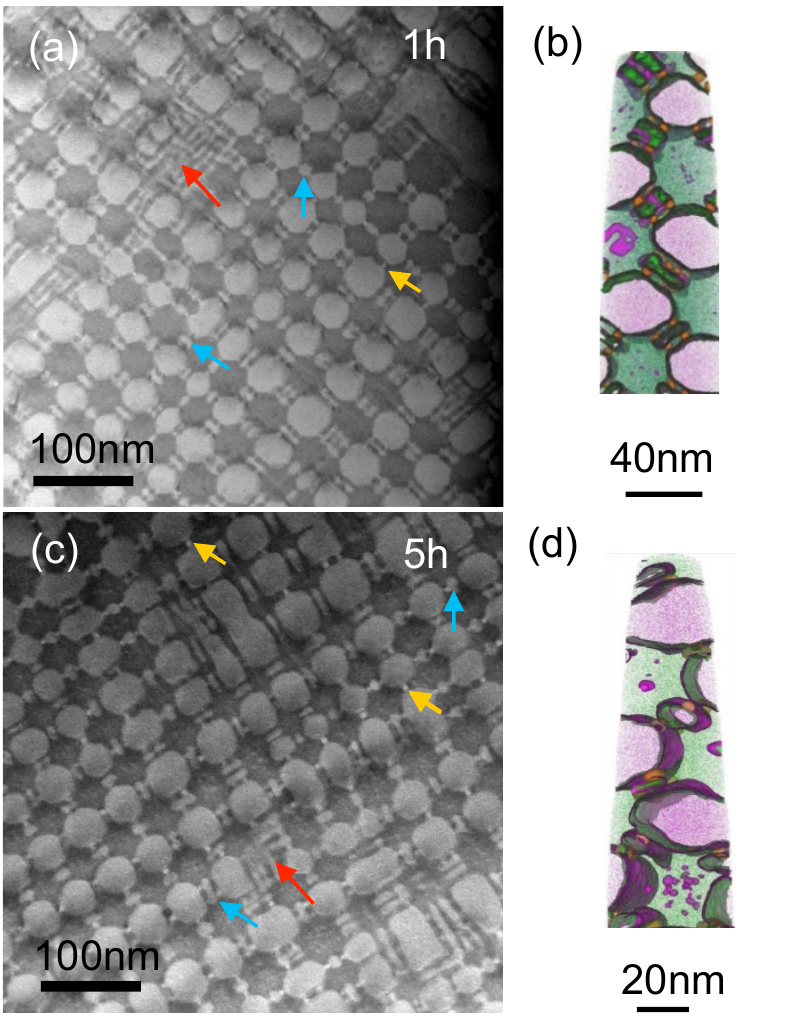}%
\caption{Nanostructures of sample 6 (a,b) and sample 8 (c,d) revealed by
HAADF STEM images and APT.  Examples of small $\ala$ rods, $\ala$
clusters and Cu-enriched rods are indicated by blue, red and yellow
arrows respectively. For clarity, the APT reconstruction only shows
the Ni (green) and Fe (pink) atoms. Isoconcentration surfaces shown
here are \SI{7.5}{\atpercent} Ni (green), \SI{10}{\atpercent} Cu
(orange) and \SI{30}{\atpercent} Fe (pink).  }
\label{fig:06}
\end{figure}

A faceted rod-shaped $\ala$ phase ($\sim$\SIrange{15}{20}{\nm}) is
developed during the MA process. As shown in \rfig{fig:04}a, a
well-defined mosaic structure composed by a $\{110\}$ faceted $\ala$
phase with a $\sim$\SIrange{10}{20}{\nm} diameter was quickly
developed in a large volume fraction of sample 2
after \SI{30}{\second} of MA. Some areas of sample 2 shows blurred
imaging contrast, as indicated by white arrows, which implies regions
with morphological differences. Increasing the MA time
to \SI{90}{\second} (sample 3, \rfig{fig:04}c) caused a slight
increase in the $\ala$ diameter to
$\sim$\SIrange{20}{30}{\nm}. $\{100\}$ facets started to appear in
some $\ala$ phases. Small $\ala$ particles ($\sim$\SI{5}{\nm}) located
between $\{100\}$ facets of two adjacent large $\ala$ rods were also
observed, as indicated by blue arrows in \rfig{fig:04}c. Moreover,
clusters of $\ala$ particles ($\sim$\SI{5}{\nm}), as pointed out by
red arrows, were formed, possibly from the white arrow regions
indicated in \rfig{fig:04}a. A further MA time increase
to \SI{10}{\minute} (sample 5, \rfig{fig:04}e) modified the $\ala$
phase diameter size distribution into a bimodal distribution with
large ($\sim$\SIrange{25}{40}{\nm}) and small ($\sim$\SI{5}{\nm})
$\ala$ phases. Isoconcentration surfaces within the APT data clearly
show $\ala$ phase elongation in all MA-treated samples, as
demonstrated in \rfig{fig:04}b, d, and f. Regions with a morphology
similar to sample 1 is also visible, as indicated by the black arrow
in
\rfig{fig:04}b, which may be regions indicated by the white arrow in
\rfig{fig:04}a. These regions are most likely areas from the solutionization
process that have not yet been modified from short time MA
processes. Transformation of big isolated $\ala$ blocks from a
parallelogram shape into an octagon shape with a cross-sectional
diameter of $\sim$\SI{35}{\nm} after \SI{10}{\minute} MA is also
obvious. For all samples, the $\alb$ phase is continuous.

Isoconcentration surfaces within the APT data reveal that the location
of Cu-enriched clusters tends to follow the edge of two adjacent
$\{110\}$ facets in sample 2 (\rfig{fig:04}b). With increasing MA
time, the region between two $\{100\}$ facets of $\ala$ shows a much
higher Cu-enriched cluster density, as shown in \rfig{fig:05}a. This
is also the area where most of the small $\ala$ phase is
located. \rFig{fig:05}b shows a HRSTEM image of sample 5. Small $\ala$
phases with a $\sim$\SIrange{3}{5}{\nm} size are clearly visible
between two $\{100\}$ facets of large $\ala$ rods. Locations of the
Cu-clusters (indicated by orange arrows) were identified by comparing
the matching EDS Cu elemental mapping (\rfig{fig:05}c). FFT analysis
(\rfig{fig:05}d) shows that the Cu/$\alb$ interface is coherent, which
implies that the Cu-clusters still have the same lattice structure as
$\alb$ after \SI{10}{\min} of MA. APT data shows no obvious change in
Cu-cluster diameter ($\sim$\SIrange{1}{1.3}{\nm}), volume fraction
(\SIrange{0.8}{1.1}{\percent}) and composition
(\SIrange{55}{68}{\atpercent}) from sample 2 thru 5.

\subsection{Microstructure after low temperature drawing }

\begin{figure}[!htb]
\centering
\includegraphics[width=\wlfig,clip]{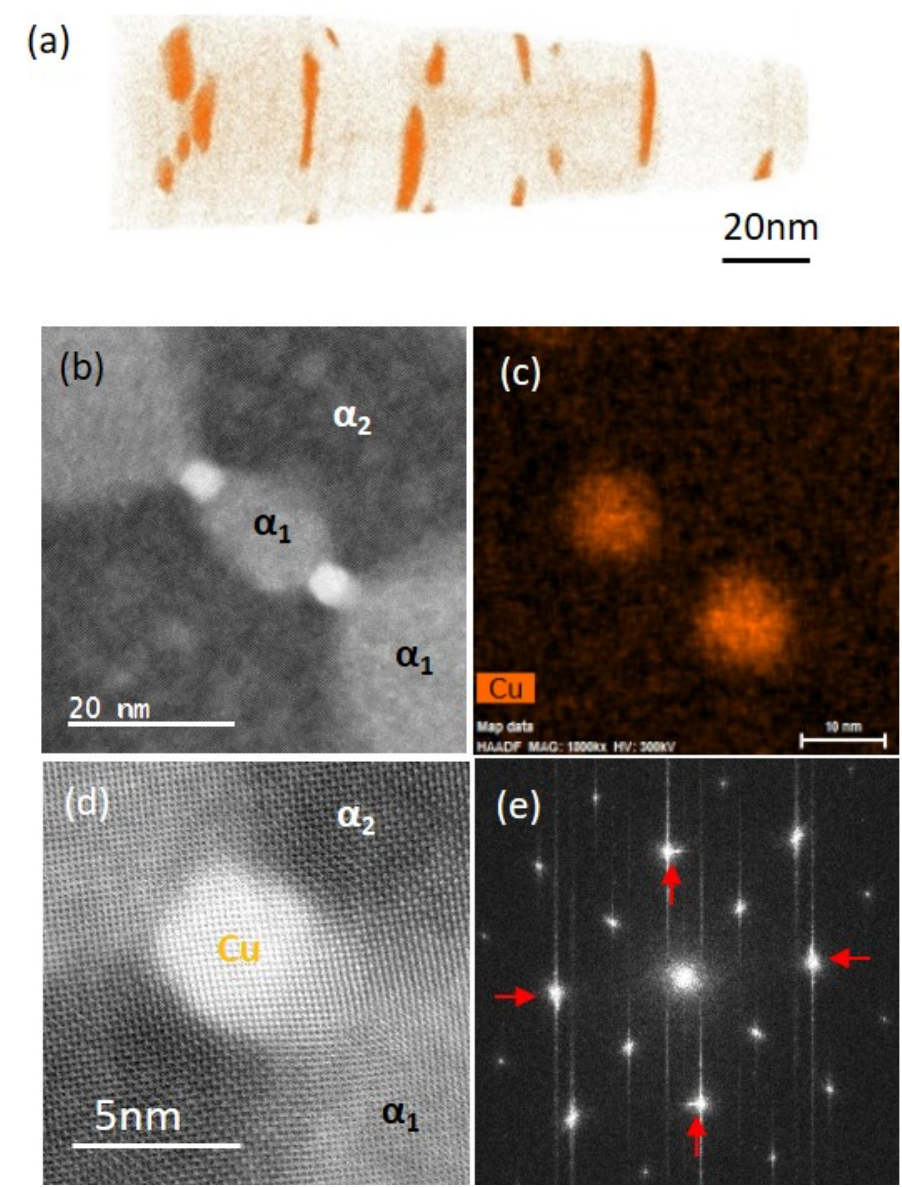}%
\caption{Sample 7 (a) Cu atom map in the reconstructed APT volume,
(b) HAADF STEM image and corresponding (c) EDS Cu mapping (scale bar
is \SI{10}{\nm}). (d) HRSTEM image shows lattice distortion in the
Cu-enriched phase. (e) FFT of (d) with red arrows indicating splitting
of (110) diffraction spots.}
\label{fig:07}
\end{figure}

\begin{figure}[htb]
\centering
\includegraphics[width=\wlfig,clip]{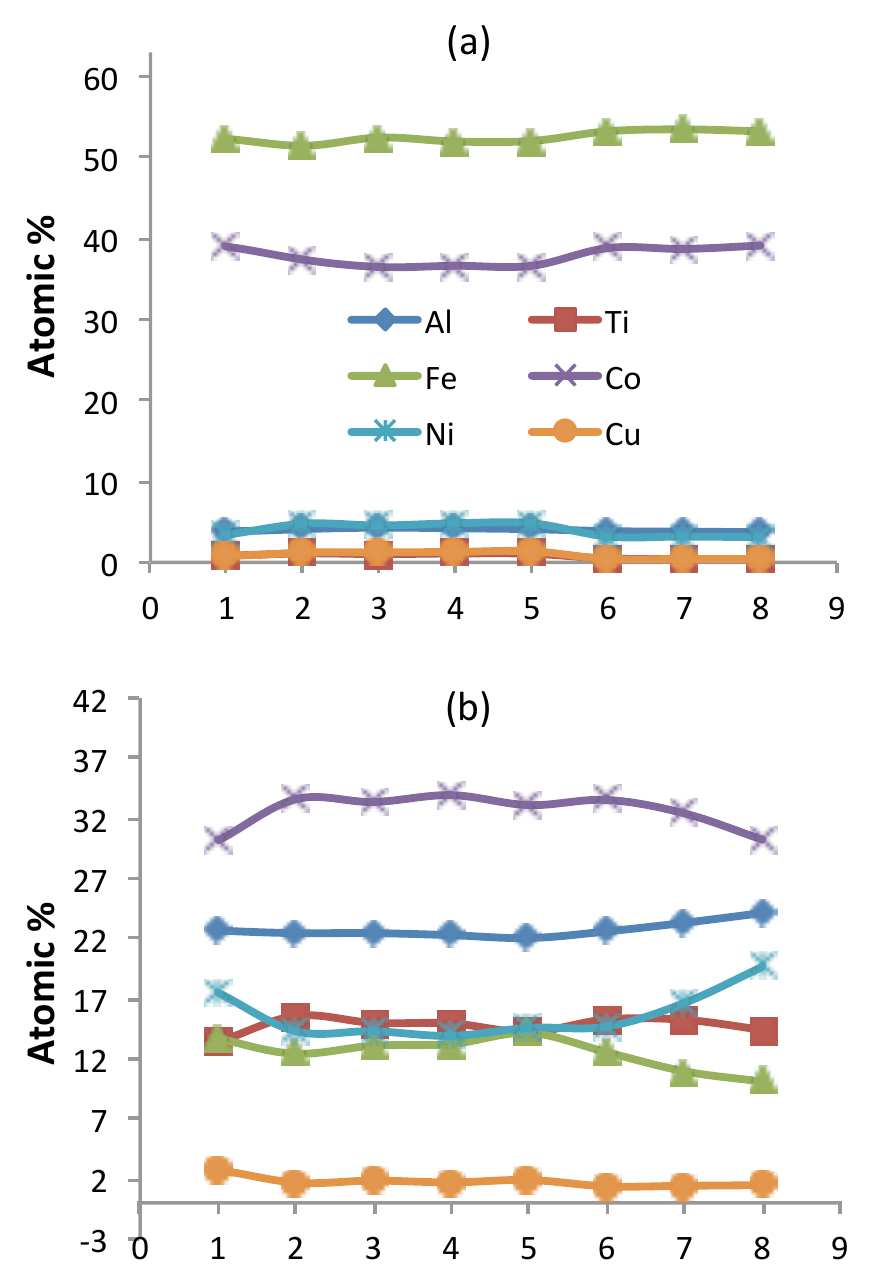}%
\caption{The elemental composition of $\ala$
(a) and $\alb$ phases (b) measured by atom probe tomography from sample 1
thru 8.}
\label{fig:08}
\end{figure}

A slight increase of large $\ala$ rod diameters to
$\sim$\SIrange{35}{60}{\nm} after a \SI{650}{\celsius} low temperature
draw was observed by STEM imaging and APT, as shown in
\rfig{fig:06}. Small $\ala$ phases in between two large $\ala$ phases agglomerate and
form smaller $\ala$ rods with a $\sim$\SI{15}{\nm} diameter, as
indicated by blue arrows in \rfig{fig:06}a and c. Regions with $\ala$
clusters, as indicated by red arrows in \rfig{fig:06}a, tend to
disappear after longer drawing hours (\rfig{fig:06}c). Moreover,
additional $\ala$ phase precipitates from the $\alb$ phase after
longer drawing hours, as shown by the pink isosurfaces within the
$\alb$ phases in \rfig{fig:06}b and d. These $\ala$ clusters are
slightly bigger in sample 8 than in sample 6.

A distinctive morphological change was observed for the Cu-enriched
phase after drawing. A transformation from clusters to rod shapes
occurs, as shown by the APT Cu elemental map in \rfig{fig:07}a. From
samples 6 to 8, the Cu-rods show an average diameter of
$\sim$\SI{3.3}{\nm}, which is 2-3 times larger than those found in
sample 5. Moreover, the central composition of Cu tends to increase
substantially from \SI{60}{\atpercent} in sample 5
to \SI{82}{\atpercent} in sample 6, and finally
to \SI{96}{\atpercent} in sample 8.  Although APT aberrations could
distort the particle concentrations, with a greater distortion for
smaller particles, structural differences revealed by STEM indicate a
higher Cu concentration (brighter contrast) for the larger particles
in sample 8, consistent with the APT results. \rFig{fig:07}b shows a
HAADF STEM image of sample 8, the bright contrast Cu-enriched phase is
clearly visible, as confirmed by matching EDS Cu elemental mapping
in \rfig{fig:07}c. The HRSTEM image shows lattice distortion in the
bright Cu-enriched phase region (\rfig{fig:07}d), which is also
manifested by a streaking/satellite peak in the (110) spots of the
image FFT (\rfig{fig:07}e), which is further evidence that the Cu
content is higher in the drawn samples.

\subsection{Chemistry evolution in $\ala$ and $\alb$ phases}
The $\ala$ and $\alb$ phase compositions in samples 1 thru 8 are summarized in
\rfig{fig:08}. Results were extracted from cropped volumes of the APT data
that were completely contained within the phases and away from the
interface, so that the chemical variation caused by the interfacial
profile could be excluded. The $\ala$ phase chemistry was relatively
stable during the whole heat-treatment process, except for a slight
increase in the Fe and Co content after drawing. More obvious
chemistry changes were detected in the $\alb$ phase. After the first
\SI{30}{\second} of MA, there was an increase of Co and Ti content, and a decrease
of Al content, while the chemical composition tends to be stable
during the following MA processes. Drawing at \SI{650}{\celsius}
gradually increased the Al and Ni concentration, while that of Co and
Fe decreased. The Ti concentration plateaued during the MA and drawing
step.  This indicates that the diffusion speed of all elements is fast
enough at \SI{840}{\celsius} to approach its thermodynamically stable
concentration within \SI{30}{\second}, however, there is an obvious
diffusion speed decrease at \SI{650}{\celsius}, and therefore, it
takes a much longer time to reach the equilibrium concentration.

\section{Discussion}
The APT combined with detailed TEM and magnetization measurements
provide a clear picture of relationships between the phase evolution
and magnetic properties. Observation of phase separation in sample 1
indicates that oil quenching cannot provide a fast enough quench to
bypass initiation of the SD. After only \SI{30}{\second} of MA, the
optimal coercivity imparted from this step is observed, even though
the APT and TEM shows further morphological changes up
to \SI{10}{\minute} of MA. The small changes in chemistry from sample
2 to 5 implies that the chemical diffusion is rapid and the observed
gradual morphology change is more likely to be driven by minimization
of interfacial and magnetic energy. After MA, particles are not only
elongated, but also become less interconnected to each other, which
helps increase $\hci$.

The \SI{650}{\celsius} draw has a remarkable effect on both $\hci$ and
refinement of the chemistry between the $\ala$ and $\alb$ phases. The
conventional explanation is that drawing further increases the
chemical separation of the two phases~\cite{mccurrie1980itom}.
However, our results suggest that the draw effect on $\hci$
enhancement likely involves several mechanisms, including chemistry,
ordering, as well as subtle structural features, such as evolution of
the Cu-enriched phase. For the $\ala$ phase, the Fe and Co
concentration slightly increases after the first draw step (comparing
sample 5 and 6), then it is nearly constant. On the other hand, in the
$\alb$ phase, Fe and Co contents continue to decrease with increasing
draw time, which is likely due to the coarsening/growth $\ala$ in
$\alb$ of precipitates. Overall, the chemical variation is small and
may not change the magnetization of the $\alb$ phase significantly. On
the other hand, site ordering in the $\alb$ phase may play an
important role. Our previous study showed that formation energy is
lower and $\tc$ decreases with increasing $\alb$ site ordering (from
BCC to DO$_3$ and L2$_1$)~\cite{zhou2014am}. Thus, draw annealing may
promote site ordering in the $\alb$ region, possibly due to a
decreases in the Fe and Co content in the $\alb$ phase. Considering
that $\tc$ of $\alb$ is near room temperature, both chemistry changes
and site ordering decrease magnetization of the $\alb$ phase at room
temperature, which increases $\hci$.

The most dramatic change is in the growth of the Cu-enriched
regions. Transformation of small Cu-enriched clusters into larger and
longer Cu-enrich rods may be driven by minimization of interfacial
energy as the center of the clusters has higher and higher Cu
concentration. Similar elongated precipitates along elastic soft
$[100]$ direction has been reported in Cu2at.\%Co system
~\cite{heinrich2007sia}. These elongated large Cu-rods may provide
better pinning for magnetic domain wall movement. The Cu-enriched
phase not only becomes bigger and longer but also less magnetic, which
can further separate $\ala$ rods. The Cu lattice shearing from the
\emph{bcc} structure may be because the \emph{fcc} structure of Cu is
thermodynamically more stable. Moreover, since some branching types
may be very detrimental to $\hci$, a larger Cu cluster can isolate two
originally connected $\ala$ rods and increases
$\hci$~\cite{ke2017apl}. Finally, formation of small $\ala$ rods or
even chains of spheres, along with the previously reported formation
of Ni-rich ($\alpha_3$) separation phase at the $\ala$/ $\alb$
interface can also help to increase $\hci$~\cite{nguyen2017pra}.

\section{Conclusions}
Coercivity enhancement is a complex interplay between the intrinsic
properties of the alnico alloy and its nanostructure. With MA, the
kinetics of the SD is rapid and the near optimum geometric spacing is
quickly reached due to higher annealing temperature. MA sets the
template for the spinodal and locks in remanence, while the draw
process is responsible for the finer microstructural and chemical
tuning, which controls the coercivity. The profound effect of draw on
improving $\hci$ is likely due to a combination of several mechanisms,
including chemical, site ordering, and subtle microstructural
variations. The draw process does not introduce dramatic
microstructural changes of $\ala$ and $\alb$ phases, but does affect
the size, shape, and distribution of the intervening Cu-rich phase
forming in-between these phases. This new understanding provides
possible directions for further property enhancement of alnico.

\section*{Acknowledgment}
Research was supported by U.S. DOE, Office of Energy Efficiency and
Renewable Energy (EERE), under its Vehicle Technologies Office,
Electric Drive Technology Program, through the Ames Laboratory, Iowa
State University under contract DE-AC02-07CH11358. APT rwas conducted
at ORNL's Center for Nanophase Materials Sciences (CNMS), which is a
DOE Office of Science User Facility.

\section*{References}

\bibliography{../../../../refs/bib/references_smpl.bib}

\end{document}